\documentclass[aps,prb,preprint,showpacs,amsmath,amssymb,unsortedaddress]{revtex4-1}

\usepackage{graphicx}
\usepackage{ulem}

\begin{document}

%\preprint{}

\title{Experimental Evidence of Cage Conduction Bands in Superconducting Cement 12CaO$\cdot$7Al$_2$O$_3$}

% authors listed in desired order
% hence multiple definitions of affiliations
\author{J. A. McLeod}
%\author{A. Moewes}
\affiliation{Department of Physics and Engineering Physics, University of Saskatchewan, 116 Science Place, Saskatoon, Saskatchewan S7N 5E2, Canada}
\email[Contact Author:]{john.mcleod@usask.ca}

\author{A. Buling}
%\author{M. A. Neumann}
\affiliation{Department of Physics, University of Osnabr\"{u}ck, Barbarastr. 7, D-49069 Osnabr\"{u}ck, Germany}

\author{E. Z. Kurmaev}
%\author{L. D. Finkelstein}
\affiliation{Institute of Metal Physics, Russian Academy of Sciences-Ural Division, 620990 Yekaterinburg, Russia}

\author{P. V. Sushko}
\affiliation{Department of Physics and Astronomy and London Centre for Nanotechnology, University College London, Gower Street, London WC1E 6BT, United Kingdom. WPI-Advanced Institute for Materials Research, Tohoku University, 2-1-1, Katahira, Aoba-ku, Sendai 980-8577, Japan.}

\author{M. Neumann}
\affiliation{Department of Physics, University of Osnabr\"{u}ck, Barbarastr. 7, D-49069 Osnabr\"{u}ck, Germany}

\author{L. D. Finkelstein}
\affiliation{Institute of Metal Physics, Russian Academy of Sciences-Ural Division, 620219 Yekaterinburg, Russia}

%\author{Yu. A. Izyumov}
%\affiliation{Institute of Metal Physics, Russian Academy of Sciences-Ural Division, 620219 Yekaterinburg, Russia}

\author{S.-W. Kim}
%\author{H. Hosono}
\affiliation{Materials and Structures Laboratory, Tokyo Institute of Technology, 4259 Nagatsuta, Midori-ku, Yokohama 226-8503, Japan}

\author{H. Hosono}
\affiliation{Materials and Structures Laboratory, Tokyo Institute of Technology, 4259 Nagatsuta, Midori-ku, Yokohama 226-8503, Japan}

\author{A. Moewes}
\affiliation{Department of Physics and Engineering Physics, University of Saskatchewan, 116 Science Place, Saskatoon, Saskatchewan S7N 5E2, Canada}
\date{\today}

\begin{abstract}
Natural 12CaO$\cdot$7Al$_2$O$_3$ (C12A7) is a wide bandgap insulator, but conductivity can be realized by introducing oxygen deficiency. Currently, there are two competing models explaining conductivity in oxygen-deficient C12A7, one involving the electron transfer via  a ``cage conduction band'' inside the nominal band gap, the other involving electron hopping along framework lattice sites. To help resolve this debate, we probe insulating and conducting C12A7 with X-ray emission, X-ray absorption, and X-ray photoemission spectroscopy, which provide a full picture of both the valence and conduction band edges in these materials. These measurements suggest the existence of a narrow conduction band between the main conduction and valence bands common in both conducting and insulating C12A7 and support the theory that free electrons in oxygen-deficient C12A7 occupy the low-energy states of this narrow band. Our measurements are corroborated with density functional theory calculations.
\end{abstract}

%\pacs{}
%\keywords{}

\maketitle

\section{Introduction}

The nanoporous complex oxide 12CaO$\cdot$7Al$_2$O$_3$ (C12A7), commonly known as the mineral mayenite, is produced on a large-scale for use in industry as a main constituent of commercial alumina cements. Although at present C12A7 is mainly used as a structural material, it has huge potential for other applications, in particular in electronics~\cite{hayashi_02b} and catalysis,~\cite{Haritha_2007_pinacol,Sojka_2008_N2O} if its unusual properties are fully utilized. C12A7 has already been investigated as a candidate material for organic light-emitting diodes,~\cite{kim_07b} resistive random access memory,~\cite{adachi_09} and as a substrate for nano-scale circuits~\cite{nishio_08}, and many more applications are possible.~\cite{kamiya_05}

\begin{figure}
\includegraphics[width=3in]{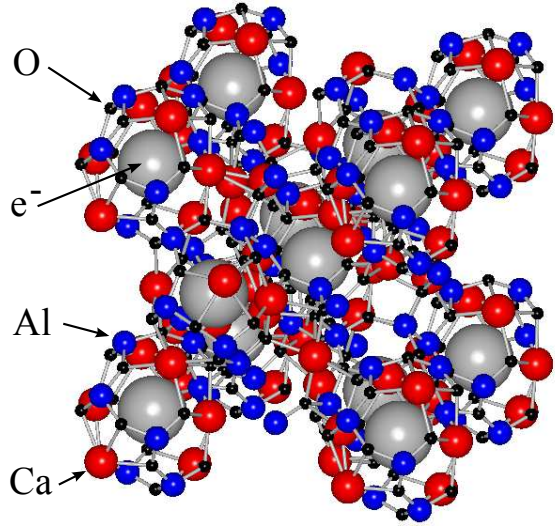}
\caption{(Color online) The lattice structure of C12A7:e$^-$ (the location of the electrons is given for visualization purpose only). The lattice is a body centered cubic (in the \textit{I}$\bar{\textit{4}}$\textit{3d} spacegroup) with a lattice constant of 11.989 \AA.~\cite{hosono_87}} 
\label{structure}
\end{figure}

The unit cell of C12A7 consists of two components and can be represented using the formula: [Ca$_{24}$Al$_{28}$O$_{64}$]$^{4+} \, \cdot$(O$^{2-}$)$_2$. Here [Ca$_{24}$Al$_{28}$O$_{64}$]$^{4+}$ denotes a three-dimensional lattice framework containing 12 cages, as shown in Figure \ref{structure}, and the two O$^{2-}$ ions refer to extra-framework species occupying some of the framework cages. These extra-framework oxide ions are more loosely bound to the lattice than the framework ones and, therefore, they are often called ``free" O$^{2-}$ ions. These species  can be partially or completely replaced by various monovalent anions, such as OH$^-$,~\cite{jeevaratnam_64} F$^-$, and Cl$^-$.~\cite{imlach_71} The replacement of the free oxide ions with O$^-$ and H$^-$ ions gives the material novel functionality, such as strong oxidation power~\cite{hayashi_02} and persistent light-induced electronic conduction,~\cite{hayashi_02b} respectively. Recently, it was reported that stoichiometric C12A7 can be converted to an n-type conductor by replacing the free oxygens with electrons using chemical reactions with Ca or Ti metals.~\cite{matsuishi_03} The extra-framework electrons in C12A7 are thought to occupy the framework cages, i.e. the lattice anion site. This electronic configuration is denoted as C12A7:e$^-$. 

With electrons replacing the free oxygens, the conductivity at room temperature is increased from 10$^{-10}$ S cm$^{-1}$ to 1.5 $\times$ 10$^3$ S cm$^{-1}$ as the electron concentration is increased~\cite{hosono_09} to 2 $\times$ 10$^{21}$ cm$^{-3}$ and a clear metal-insulator transition is observed around electron concentration of 1$\times$10$^{21}$ cm$^{-3}$.~\cite{kim_07} When the metallic conducting C12A7:e$^-$ is cooled down to low temperature, a sharp drop of resistivity takes place and a superconducting state is observed as manifested by the measurement of zero resistivity, magnetic field-dependent resistance and perfect diamagnetism. The value of T$_c$ of C12A7:e$^-$ varies in the range of 0.14-0.4 K depending on the concentration of the electron anions.~\cite{miyakawa_07}

Two different models of electronic structure for electron doped C12A7 have been suggested based on density functional theory (DFT) calculations. According to one of them~\cite{PVS_2003_C12A7_PRL,sushko_03,sushko_07} C12A7 has two conduction bands (CB): the upper CB is due to cation sites of the framework (the ``framework conduction band'', or FCB), whereas the lower and narrower CB is formed by the states associated with the cages (the ``cage conduction band'', or CCB). When electrons substitute the extra-framework anions, they occupy some of the CCB states. At low  concentrations, extra-framework electrons form polarons, which give rise to electronic conduction via thermally activated cage-to-cage hops. At high concentrations, the electronic states of extra-framework electron overlap and make the polaronic distortions smaller, which gives rise to the metallic conduction.

The alternative model suggests that extra-framework electrons hop along pathways formed by framework Ca atoms but not by the empty cages.~\cite{medvedeva_04} In this model, the extra-framework electrons are not localized in the cages but distributed over the lattice. These electrons occupy a hybrid subband at the bottom of the conduction band. To resolve the debate between these two models, it is crucial to determine the electronic structure of C12A7 by direct spectroscopic measurements. In the present work we perform a full X-ray spectroscopic study of undoped and electron doped C12A7 crystals using the techniques of X-ray photoemission spectroscopy (XPS), X-ray emission spectroscopy (XES), and X-ray absorption spectroscopy (XAS) and compare our measurements with our electronic structure calculations.

\section{Experimental and theoretical Methods}

Insulating 12CaO$\cdot$7Al$_2$O$_3$ single-crystals were grown by a FZ (Floating Zone) method. The grown single-crystals (hereafter referred to as ``Sample A'') were transparent and insulating. The details of the growth procedure have been reported earlier.~\cite{yoon_08} Conducting C12A7:e$^-$ single-crystals (hereafter referred to as ``Sample B'') were prepared by heat treatment of Sample A under Ti metal vapor. In brief, some insulating single-crystals were sealed in a silica glass tube with Ti metal powders under a vacuum. The electron concentration ($N_e$) was controlled by adjusting the temperature and duration of the heat treatment. Sample B was prepared by the heat treatment at 1100$^\circ$ C for 24 hr. The details of this procedure can be found elsewhere.~\cite{kim_07}
The electron concentration and electrical conductivity of sample B was found to be $\sim$2 $\times 10^{21}$ cm$^{-3}$, and $\sim$1000 S/cm, respectively.

The soft X-ray emission measurements of C12A7 samples were performed at the soft X-ray fluorescence endstation at Beamline 8.0.1 of the Advanced Light Source at Lawrence Berkeley National Laboratory.~\cite{jia_95} The endstation uses a Rowland circle geometry X-ray spectrometer with spherical gratings and an area sensitive multichannel detector. We have measured resonant and non-resonant O $K$ ($2p \rightarrow 1s$ transition) and Ca $L_{2,3}$ ($3d4s \rightarrow 2p$ transition), and non-resonant Al $L_{2,3}$ ($3s \rightarrow 2p$ transition) XES. The instrumental resolving power (E/$\Delta$E) for all spectra was about 10$^3$. Since we did not have the ability to cleave the samples \textit{in situ} for the XES measurements, additional O $K$ XES measurements were performed for commercial (Alfa Aesar, 99.9\% purity) TiO$_2$, CaO, and CaCO$_3$ powders pressed into clean indium foil. These measurements were to help demonstrate that our XES measurements were indeed probing the bulk C12A7 electronic states.

The O \textit{1s} and Ca \textit{2p} XAS measurements were performed at the spherical grating monochromator (SGM) beamline at the Canadian Light Source (CLS).~\cite{regier_07} The Al \textit{2p} XAS measurements were performed at the variable line spacing plane grating monochromator (VLS PGM) beamline at the CLS.~\cite{hu_07} In both cases the spectra were measured in total electron yield (TEY) and total fluorescence yield (TFY) mode, and normalized to the incident photon current on a highly transparent mesh in front of the sample to correct for intensity fluctuations in the photon beam (a gold mesh for the SGM, a nickel mesh for the VLS PGM). The XAS spectra reported here were all collected in the bulk sensitive TFY mode.

XPS measurements of the O \textit{1s}, Ca \textit{2p}, and Al \textit{2p} core levels as well as the valence band were obtained using a Perkin-Elmer PHI 5600 ci Multitechnique System with monochromatized Al \textit{K}$\alpha$ radiation with a full width at half-maximum (FWHM) of 0.3 eV. The energy resolution of the spherical capacitor analyzer was adjusted to approximately $\Delta$E = 0.45 eV. The pressure in the ultra-high vacuum chamber was in the 10$^{-10}$ mbar range during the measurements. For our XPS measurements we were able to cleave the C12A7 samples \textit{in situ}, the XPS measurements are therefore mostly free of contamination.

\begin {figure}
\includegraphics[width=3in]{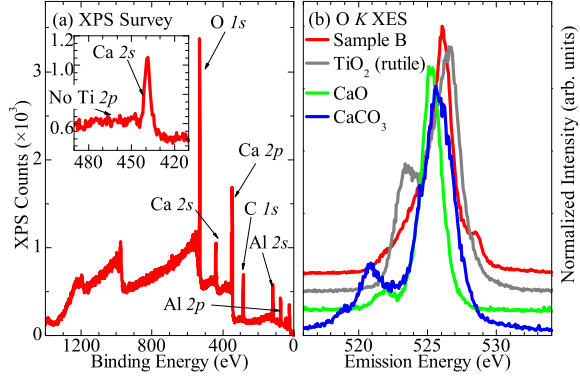}
\caption {(Color online) The survey XPS spectrum and the O \textit{K} XES spectrum of sample B (conducting). The O \textit{K} XES spectrum of sample B is compared to the O \textit{K} XES spectrum of several reference materials.}
\label{survey}
\end{figure}

We can be fairly confident of the purity of our samples. As previously mentioned, these samples were prepared in a  Ti vapour. However, the XPS survey scan of sample B (see Figure \ref{survey}) shows that the cleaved sample has no trace of Ti - confirming that the Ti deposited by the vapour treating process is confined to the surface. There is, however, a slight carbon contamination found even in the bulk of this sample. Fortunately, the O \textit{K} XES of sample B bears little resemblance to the O \textit{K} XES of TiO$_2$ (rutile), CaO, or CaCO$_3$ (see Figure \ref{survey}), indicating that neither the carbon contamination nor any potential oxide or titanide  surface layer is distorting our measurements.

The electronic structure calculations were performed using the density functional theory (DFT) and two density functionals. The geometrical structures of stoichiometric C12A7 (C12A7:O$^{2-}$) and C12A7 in which all extra-framework oxide ions are replaced by electrons (C12A7:e$^-$), were optimized using the Perdew-Burke-Ernzerhof (PBE) density functional,~\cite{PBE_1996_PRL} a plane-wave basis set, and the projected augmented waves method~\cite{Blochl_1994_PRB_PAW,Kresse_1999_PRB_VASP_PAW} implemented in the Vienna {\it Ab Initio} Simulation Package (VASP).~\cite{Kresse_1996_PRB_VASP} The plane-wave basis set cutoff was set to 500 eV. Cubic supercells containing 118 atoms, in the case stoichiometric C12A7, and 116 atoms, in the case of C12A7:e$^-$, and reciprocal space grid of eight Monkhorst-Pack $k$-points were used. In both cases the total energy of the system was minimized with respect to the atomic positions until the maximum force acting on any individual atom did not exceed 0.01 eV/\AA. The lattice constant was fixed at the experimental value of 11.989 \AA.~\cite{hosono_87} For the analysis of the electronic structure, the charge density was decomposed over atom-centered spherical harmonics. 

\section{Results and Discussion}

\begin {figure}
\includegraphics[width=3in]{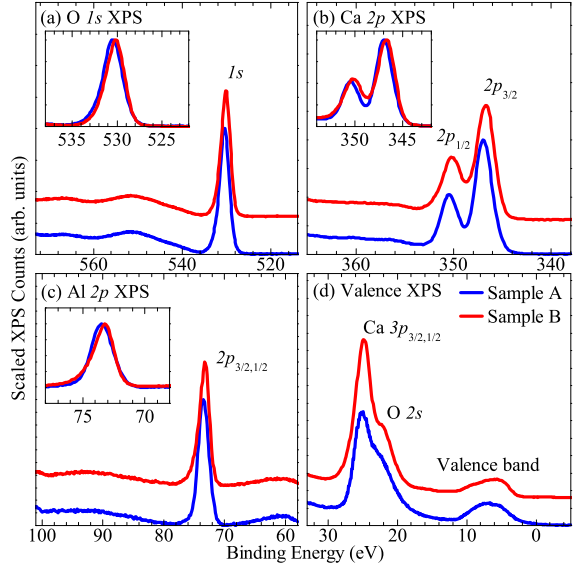}
\caption {(Color online) The XPS spectra collected for samples A and B. (a) The O \textit{1s} spectra, the inset offers a more direct comparison of the O \textit{1s} peak. (b) The Ca \textit{2p}$_{3/2}$ and \textit{2p}$_{1/2}$ spectra. (c) The Al \textit{2p}$_{3/2,1/2}$ spectra. (d) The valence band spectra, as well as the semi-core O \textit{2s} and Ca \textit{3p} peaks. }
\label{xps}
\end{figure}

The XPS spectra obtained for the three C12A7 samples are shown in Figure \ref{xps}. The crystal structure of insulating C12A7 has 3 different oxygen sites: two in the framework and one, partially occupied site, in the middle of the cages (the extra-framework or ``free" O$^{2-}$ ion site). The two types of the framework O atoms have different local atomic structures: one is surrounded by two Ca$^{2+}$ ions and two Al$^{3+}$ ions, while the other by three Ca$^{2+}$ ions and one Al$^{3+}$ ion. The width of the O \textit{1s} core-level XPS (see Figure \ref{xps}(a)) is nearly the same for all samples, however. This implies that the \textit{1s} binding energies of all three oxygen sites are the same within the resolution of our measurement. One might expect that the O$^{2-}$ site would have a sufficiently different binding energy from the lattice oxygen to be resolved in the XPS measurement, unfortunately it seems that the binding energy of the O$^{2-}$ \textit{1s} core level is too close to that of the lattice site and/or the concentration of O$^{2-}$ atoms is too small (even in sample A only 2 out of 64 oxygens are the free O$^{2-}$ ions) for this to be resolvable in our measurements.

Another factor affecting the sharpness of the core-level XPS features has to do with the ease of cleaving and crystallinity of the cleaved samples. It has been shown theoretically,~\cite{PVS_C12A7_surf} that the occupied and the empty cages behave as ``tough" and ``soft" lattice elements, respectively. Consequently, the cages occupied by extra-framework O$^{2-}$ ions remain intact during a simulated material rupture, while the empty cages break. This suggests that in sample B, where all cages are equivalent, the lack of such soft elements leads to a higher degree of disorder in the near surface region, or after forcibly cleaving a new surface region, than in samples containing extra-framework O$^{2-}$ ions (see Ref.~\onlinecite{PVS_C12A7_surf} for a detailed discussion). Since calcium forms the bulk of the cage structure, the Ca \textit{2p} XPS (see Figure \ref{xps}(b)) of sample B is broader and less well resolved than that from sample A because of the softer lattice.

Finally, because sample A is insulating, the valence XPS (see figure \ref{xps}(d)) is affected by artifacts due to charging. Because of the differences in conductivity and lattice strength, one should not read too much into the relative intensities of various core-level XPS features between different samples.

\begin {figure}
\includegraphics[width=3in]{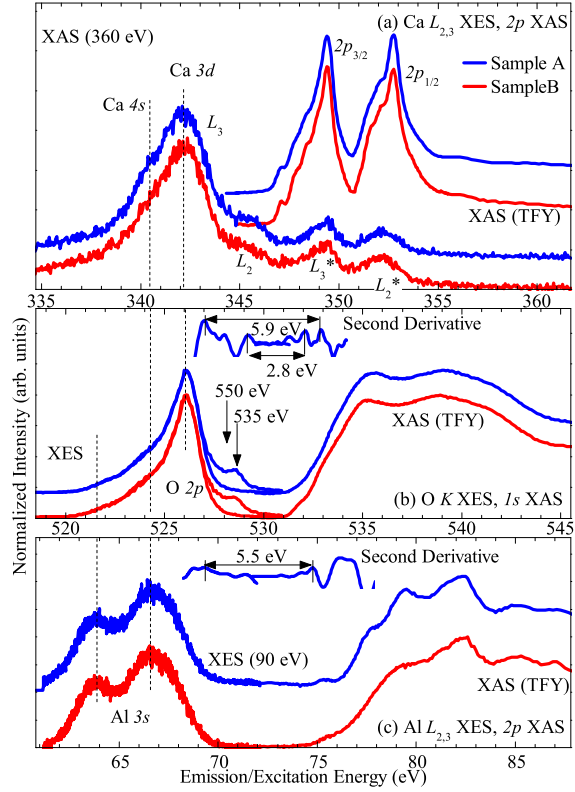}
\caption {(Color online) The XES and XAS spectra. (a) The Ca \textit{L}$_{2,3}$ XES and \textit{2p} XAS. (b) The O \textit{K} XES and \textit{1s} XAS, note that two different XES spectra corresponding to different excitation energies (535 and 550 eV) are plotted for each sample. The second derivative of the XES (excited at 535 eV) and XAS for sample A are shown in inset in the middle, on a consistent energy scale. (c) The Al \textit{L}$_{2,3}$ XES and \textit{2p} XAS, the second derivative of the XES and XAS for sample A is inset in the middle, again on a consistent same energy scale. In all panels, vertical lines emphasise the hybridization between the spectra.}
\label{xesxas}
\end{figure}

The XES and XAS measurements for the three C12A7 samples are shown in Figure \ref{xesxas}. The Ca \textit{L}$_{2,3}$ XES consists of 4 subbands. The first two subbands (shown in Figure \ref{xesxas}(a)), correspond to the Ca \textit{L}$_3$ XES (the \textit{3d4s}$\rightarrow$\textit{2p}$_{3/2}$ transition) and Ca \textit{L}$_2$ XES (the \textit{3d4s}$\rightarrow$\textit{2p}$_{1/2}$ transition), respectively. The next two subbands in the Ca \textit{L}$_{2,3}$ XES (denoted as \textit{L}$_3^*$ and \textit{L}$_2^*$) coincide in energy with the Ca \textit{2p} XAS absorption peaks. These are due to a relaxation of the photo-excited electrons into the long lifetimes states above the Fermi energy, which then undergo a second fluorescence transition to the Ca \textit{2p} core level.  This is the process of re-emission,~\cite{yablonskikh_01} it is known to occur in the X-ray emission of elements that formally have no valence electrons.~\cite{fomichev_68} The core-hole created by the photo-excitation process causes a local perturbation in which a \textit{2p}$^5$(\textit{e},\textit{t}$_2$)$^1$ $\rightarrow$ 2p$^6$(\textit{e},\textit{t}$_2$)$^0$ transition occurs.~\footnote{Note that because the X-ray induced transition temporarily breaks crystal symmetry, the \textit{e} and \textit{t}$_2$ levels referred to here are only local to the excited atom, and do not refer to the character of the crystal band structure.} This perturbative transition allows a photo-excited electron decay from a normally unoccupied level.~\cite{curelaru_79} While this effect is physically interesting, it does not give much information about the ground state electronic structure of the material.

The valence band in C12A7 is dominated by O \textit{2p} states. These are probed by O \textit{K} XES measurements (the \textit{2p} $\rightarrow$ \textit{1s} transition). The corresponding spectrum is shown in Figure \ref{xesxas}(b). Note that for all C12A7 samples the non-resonant O \textit{K} XES, excited at 550 eV or higher energies, has a high energy shoulder, while resonant excitation, excited at 535 eV, removes this feature.

Previous research on simple oxides has shown using the peaks in the second derivatives of the XES and XAS spectra can provide a good estimate of the band gap.~\cite{kurmaev_08,mcLeod_10} Since the effect of the O \textit{1s} core-hole on the O \textit{1s} XAS typically causes a local increase in the number of states near the bottom of the conduction band, but does not shift the bottom of the conduction band to lower energies~\cite{mcLeod_10} as is common with a \textit{2p} core-hole in metals,~\cite{kurmaev_08} the separation between the O \textit{K} XES and \textit{1s} XAS can be used to accurately estimate the band gap. Because there is a high energy shoulder in the 550 eV O \textit{K} XES, we chose the main peak of its second derivative as the top of the framework valence band, and the peak in the second derivative near the edge of the high energy shoulder as the top of the \textit{2p} states of extra-framework O$^{2-}$ ions formed by the states associated with extra-framework O$^{2-}$ ions.  Similarly, we interpret the lower and higher energy peaks in the second derivative of the O \textit{1s} XAS (these peaks are of the same intensity) as the bottom of the cage conduction band and the framework conduction band, respectively. These choices predict a framework gap of 5.9 eV and a band gap (defined from the O$^{2-}$ valence states to the lower conduction band) of 2.8 eV (as shown in Figure \ref{xesxas}(b)).

The spin-orbit splitting in Al is much smaller than that in Ca (0.4 eV compared to 3.5 eV ~\cite{elam_02}). Since the \textit{L}$_3$ emission band is wider and the core-hole broadening is greater than 0.4 eV, the Al \textit{L}$_{2}$ is indistinguishable from the Al \textit{L}$_3$ XES (see Figure \ref{xesxas}(c)).  The Al \textit{L}$_{2,3}$ XES of C12A7 is quite similar to both the measured and theoretical Al \textit{L}$_{2,3}$ XES of crystalline and amorphous Al$_2$O$_3$.~\cite{zatsepin_04,simunek_93} Unlike the Ca \textit{2p} XAS, the Al \textit{L}$_{2,3}$ XES and \textit{2p} XAS are not strongly affected by multiplets, and therefore both can be directly compared to the ground state electronic structure. Because of this, the separation between the Al \textit{L}$_{2,3}$ XES and \textit{2p} XAS can be used to estimate the band gap. Applying the above second derivative method to the Al \textit{L}$_{2,3}$ XES and \textit{2p} XAS indicates a separation between the Al \textit{L}$_2$ XES and the Al \textit{2p}$_{3/2}$ XAS of 5.5 eV. Note that this separation is between spectra from two different core levels. Because the Al \textit{L}$_3$ XES is 0.4 eV lower than the Al \textit{L}$_2$ XES (or, alternatively, because the Al \textit{2p}$_{1/2}$ XAS is 0.4 eV higher than the Al \textit{2p}$_{3/2}$ XAS), the actual band gap is closer to 5.9 eV (denoted in Figure \ref{xesxas}(c)) - as previously predicted by the O \textit{K} XES and \textit{1s} XAS. Since the Al \textit{3s} states are not expected to be a major component of the cage conduction band, we anticipate that these measurements probe the framework band gap. Note this also implies that the Al \textit{2p} core-hole causes a negligible shift in the Al \textit{2p} XAS, as has been previously suggested.~\cite{tanaka_96}

Finally, the vertical lines in Figure \ref{xesxas} denote the hybridizations apparent in the XES spectra. The individual energy scales in Figure \ref{xesxas} have been aligned according to the peaks in the second derivative. For the Ca \textit{L}$_{2,3}$ XES, the peak in the second derivative closest to the \textit{L}$_3$ emission band was chosen for alignment. Aligning the C \textit{L}$_3$ emission with the other spectra is difficult, because the Ca \textit{L}$_2$ emission band partially overlaps the Ca \textit{L}$_3$ emission band. However the chosen peak in the second derivative near the top of the \textit{L}$_{3}$ band is roughly 3 eV lower than the peak in the second derivative at the top of the \textit{L}$_2$ band, which is close to the actual spin-orbit splitting of Ca (3.5 eV). Because the Ca \textit{L}$_2$ emission band is weak compared to the noise threshold, this method of alignment is sufficient for our purposes.

The shape of the Ca \textit{L}$_3$ emission band shows the separation of low energy \textit{s}-symmetry valence states and higher energy \textit{d}-symmetry valence states, as is common in binary oxides.~\cite{zatsepin_04} With regard to the top of the valence band, the Ca \textit{3d} emission band coincides in energy with the main peak in the O \textit{2p} states (see Figure \ref{xesxas}(a,b)). The Ca \textit{4s} emission band is the shoulder at 340.4 eV, this coincides in energy with an Al \textit{3s} emission band. There is also a shoulder in the O \textit{2p} emission band at the same relative energy, indicating Ca \textit{4s} - Al \textit{3s} - O \textit{2p} hybridization in this region. Lastly, the Al \textit{3s} emission band coincides with a weak shoulder at 521.8 eV in the O \textit{2p} emission band. 

\begin {figure}
\includegraphics[width=3in]{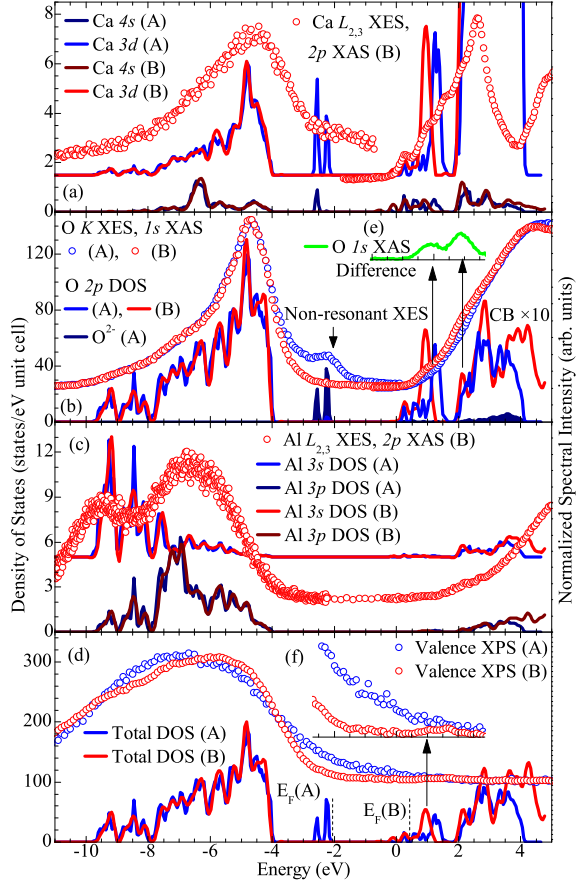}
\caption {(Color online) The calculated DOS (lines) compared  to the XES, XAS, and valence XPS measurements (circles). The DOS projected on the atomic orbitals of each atomic species are plotted along with the corresponding XES and XAS spectra. (a) The Ca states and spectra, note that the Ca \textit{L}$_2$ emission has been cut off at $\sim -0.5$ eV due to the onset of the reemission lines shown in Figure \ref{xesxas}, and the spectral weight in the Ca \textit{L}$_{2,3}$ XES between -3 and 0 eV is due to the \textit{L}$_2$ emission line, not background counts. (b) the O states and spectra (note that the non-resonantly excited O \textit{K} XES for sample A and the resonantly excited O \textit{K} XES for sample B is shown), (c) the Al states and spectra, and (d) the total valence states and spectra. The difference between the O \textit{1s} XAS of samples B and A is plotted in (e) in green in the middle of the figure, while a magnification of the valence XPS near the Fermi level is plotted in (f) at the bottom of the figure. The calculated highest occupied state in C12A7:O$^{2-}$ and the Fermi level in C12A7:e$^-$ are labeled as E$_\mathrm{F}$(A) and E$_\mathrm{F}$(B), respectively. The energy scale is aligned with the valence XPS measurements, and is accurate with respect to the core level energies of Figure \ref{xps}.}
\label{dos}
\end{figure}

These experimental results are compared with the calculated total and atom-projected densities of states (DOS) in Figure~\ref{dos}. The measured XES and XAS spectra were aligned with the valence XPS spectra based on the measured XPS core level binding energies, the calculated DOS was aligned with this energy scale based on the common hybridization features mentioned above. As it is clear from Figure~\ref{dos}, all experimentally observed features of these spectra are in excellent agreement with the results of the calculations. This is not a trivial result: The experimental spectra in Figure~\ref{xesxas} predict the relative energies of the bulk of the O \textit{2p}, Ca \textit{3d}, Ca \textit{4s}, and Al \textit{3s} states, and our alignment of the DOS with the experimental spectra only guarantees the agreement of a single feature - we can align the calculated O \textit{2p} states with the O \textit{K} XES spectra, but that does not guarantee that the calculated Ca and Al states will agree with the measured Ca and Al \textit{L}$_{2,3}$ spectra, or that any of the XAS spectra will agree with the calculated conduction band states. It is also important to point out that 0 eV on the XPS binding energy scale does not refer to the actual Fermi energy of either samples A or B; since the highest occupied state is at quite different energies in both compounds, and since the actual Fermi level of B is difficult to experimentally determine due to the relatively low number of states at this energy, the zero point for the binding energy scale was chosen based on comparison with known reference standards. This had the advantage of providing a consistent energy scale for both samples A and B.

This comparison of theory and experiment in Figure \ref{dos}(a) reveals that, as predicted above, the Ca \textit{L}$_{3}$ XES is primarily comprised of Ca \textit{3d} states, with some Ca \textit{4s} states contributing at lower energies (at -6 eV). Similarly, the Al \textit{L}$_{2,3}$ XES consists of a low energy band of Al \textit{3s} states (at -8 to -10 eV). The higher energy band in the Al \textit{L}$_{2,3}$ XES, from -4 to -8 eV, coincides in energy with the Al \textit{3p} states. Since a \textit{3p} $\rightarrow$ \textit{2p} transition is dipole forbidden, it is likely that the projected DOS underestimates the amount of Al \textit{3s}, \textit{3p}, \textit{3d} hybridization in this region. Finally, the unoccupied Al states have negligible hybridization with the cage conduction band, so our earlier prediction that the Al \textit{L}$_{2,3}$ XES and \textit{2p} XAS could be used to estimate the framework band gap is well founded.

As previously mentioned, the O \textit{K} XES probes the O \textit{2p} states, which the calculations show are hybridized primarily with Ca \textit{3d} states between -4 and -6 eV, and Al \textit{3s} states at lower energies. While the states from the O$^{2-}$ ions can be accurately represented by the high energy shoulder in the non-resonant O \textit{K} XES, there is also a known high-energy satellite line of oxygen~\cite{valjakka_85} in this range as well. Since a high energy shoulder of roughly the same intensity appears in the O \textit{K} XES of all samples (see Figure \ref{xesxas}), it is clear it cannot be due to the O$^{2-}$ alone, despite the coincidence in energy. The satellite line is due to double ionization in the valence band (a Wentzel-Druyvesteyn satellite),~\cite{druyvesteyn_27} and can be suppressed by low energy excitation. However, since the O$^{2-}$ ions have very few unoccupied states at low energies in the conduction band, as seen in the calculated partial DOS of the O$^{2-}$ atoms (Figure \ref{dos}(b)), any emission from these states will also be suppressed by low energy excitation. The unfortunate overlap of these two very different effects prevents the unambiguous detection of the O$^{2-}$ states in the XES.

The calculation also suggests some hybridization between Ca \textit{3d} states and the O$^{2-}$ \textit{2p} states, however unfortunately the Ca \textit{L}$_{2,3}$ is also unable to unambiguously resolve this feature. This is because the high energy shoulder of the Ca XES spectrum is the \textit{L}$_2$ emission line (refer back to Figure \ref{xesxas}, so it is difficult to distinguish high energy \textit{L}$_3$ emission from the main \textit{L}$_2$ emission. Secondly the appearance of hybridized Ca \textit{3d} and O$^{2-}$ \textit{2p} states in this region may easily be an artifact of the parameters used in the calculation: Some of the O$^{2-}$ valence charge will enter the region near the Ca atom, but the actual overlap of the Ca \textit{2p} core electron (near the Ca nucleus) and this valence charge (near the somewhat artificial edge of the Ca atomic sphere) is small. Since the overlap integral of the core and valence wavefunctions is responsible for the amplitude of the X-ray transition,~\cite{schwarz_75} the XES spectral intensity in this region due to this hybridization is likely quite a bit smaller than the intensity of the DOS suggests.

It is important to point out that our simple method of aligning the measurements to the calculated DOS provides excellent agreement in terms of the relative energies of the calculated and measured features. The method of calculating the band gaps using the second derivatives of the experimental spectra is in almost complete agreement with the band gaps calculated using DFT: with the top of the framework valence band at -4 eV, as shown in Figure \ref{dos}, the top of the framework conduction band is at 1.9 eV as predicted by both the Al \textit{L}$_{2,3}$ XES and \textit{2p} XAS measurements, and the O \textit{K} XES and \textit{1s} XAS measurements. We ought to add, however, that the quantitative agreement between the calculated and experimentally observed features might be fortuitous. Indeed, the one-electron band gaps obtained using GGA DFT method are known to be underestimated. On the other hand, this error might be compensated for by the interaction of the core holes and excited electrons, which tends to lower the apparent band gap and is not reflected in the DOSes. The alignment between calculation and spectra is determined within an accuracy of a few tenths of an eV, and shifting the measured spectra by that amount in energy along either direction would still give decent agreement with the calculated states.

The calculated gap between the \textit{2p} states of extra-framework O$^{2-}$ ions and the cage conduction band in C12A7:O$^{2-}$ is 2.2 eV,  which is somewhat smaller than the 2.8 eV predicted from the O \textit{K} XES and \textit{1s} XAS spectra of sample A. This could be due to the aforementioned tendency of GGA DFT to underestimate the values of band gaps. Another possible source of this disagreement is the gradual increase of states in the cage conduction band (the peak of the cage conduction band is 1.5 eV above the onset of this band). Indeed, there is a lower energy and smaller intensity peak in the second derivative of the O \textit{1s} XAS spectra (at 531.2 eV in Figure \ref{xesxas}) that, if used as an estimate of the bottom of the cage conduction band, would predict a gap between the \textit{2p} states of extra-framework O$^{2-}$ ions and cage conduction band of 2.1 eV. This suggests that for a narrow isolated conduction band, the second derivative method can identify both the lower energy and the higher energy edges of the band. Then, the separation of 0.6 eV between these peaks can be associated with the width of the cage conduction band.

Since there is only a subtle change in electronic structure between insulating and conducting C12A7, the measurements of samples A and B are almost identical. There are, however, two key differences.

First, the onset of the O \textit{1s} XAS spectrum of sample B has a greater amplitude than that of sample A. The difference between them is shown as two narrow peaks in Figure \ref{dos}(e). These two peaks are in the energy range, where the magnitude of the DOS calculated for the cage and framework conduction bands in C12A7:e$^-$ is greater than that in C12A7:O$^{2-}$. Because the core-hole effect in O \textit{1s} XAS measurements causes an increase in the amplitude of states close to the Fermi level,~\cite{mcLeod_10} we expect the somewhat minor differences in the calculated states of conducting and insulating C12A7 to be amplified in the differences between the O \textit{1s} XAS of samples B and A. 

Second, the valence XPS spectrum of sample B shows a small but statistically significant feature 5 eV above the edge of the framework valence band (shown in Figure \ref{dos}(f), as noted above the 0 eV value on the binding energy scale does not necessarily refer to the Fermi energy, so we expect the states shown in Figure \ref{dos}(f) to represent occupied states in the cage conduction band and also possibly some long-lifetime excited states). This feature is close to the energy where the occupied states of the cage conduction band in C12A7:e$^-$ are predicted to be, and this feature suggests that the calculated gap between the CCB and the framework valence band (i.e. the calculated Fermi level) is underestimated. We point out that this feature was also observed in hard X-ray valence XPS measurements of C12A7:e$^-$.~\cite{toda_07} Recent low temperature (T $\sim$ 10 K) Xe-sourced XPS measurements of C12A7:e$^-$ produced a clearer picture of the \textit{2p} states of extra-framework O$^{2-}$ ions at about 5 eV above the edge of the framework valence band.~\cite{souma_10} Unfortunately the O$^{2-}$ valence states at $\sim -2.5$ eV in sample A can not be unambiguously resolved in the valence XPS due to sample charging.

\begin {figure}
\includegraphics[width=3in]{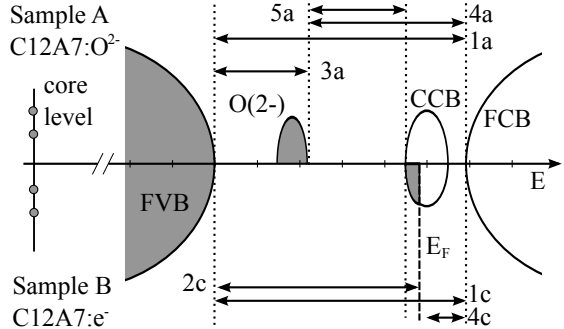}
\caption {(Color online) A stylized schematic of the C12A7 electronic structure. Occupied states are shaded while unoccupied states are clear. The Fermi level of C12A7:e$^-$ is noted, and the gaps between regions that can be determined by our spectroscopy techniques are labeled. The figure is to scale in terms of energy, each tick mark on the central energy axis represents 1 eV. (See Table \ref{tbl:gaps} for numerical values.)}
\label{transitions}
\end{figure}

To clarify the comparison between our spectroscopy measurements and calculated electronic structure, we can resort to a simpler description of the electronic states. By referring again to Figure \ref{dos}, we see that both insulating and conducting C12A7 (samples A and B, respectively) have a framework valence band (FVB) below -4 eV, and the aforementioned FCB above 1.9 eV. Sample A has \textit{2p} states of free O$^{2-}$ ions (O(2-)) near -2.5 eV, while sample B has occupied states at the bottom of the cage conduction band (CCB) near 0 eV. Finally both samples have the maximum of aforementioned CCB at roughly 1 eV. Since our spectroscopic techniques are well suited to determining the tops of valence bands and the bottoms of conduction bands, we therefore concern ourselves with the gaps described in Figure \ref{transitions}.

The calculated and measured values (from XES and XAS data) of the gaps shown in Figure \ref{transitions} are given in Table \ref{tbl:gaps}. There is good agreement between our calculated and measured gaps. Furthermore, there is good agreement between our results and those from the literature except where the FCB is concerned. In particular, in some cases our experimentally measured gaps are lower than the calculated gaps (for example, the O(2-)-FCB gap). This is unusual, given that GGA DFT is known to underestimate gaps. The culprit in this situation is the core-hole effect in the XAS measurements. Although, as previously mentioned, the core-hole is not expected to shift the energy of the onset of the conduction band in our O \textit{1s} XAS measurement, the actual onset of the conduction band is the CCB in both samples. Therefore, because the core-hole effect causes an increase in the number of states near the edge of the conduction band, the gap between the CCB and the FCB may be reduced, or even closed completely in a XAS spectrum. This is likely the reason why the framework band gap (labeled as 1a,c in Figure \ref{transitions}) is smaller in our measurements than in optical measurements. Additionally, the discrepancy between the measured and calculated O(2-)-CCB gap and the FVB-O(2-) gap is likely to occur because the onset of the extra-framework O$^{2-}$ states is obscured by the above-mentioned Wentzel-Druyvesteyn satellite in the non-resonant O \textit{K} XES.

\begin{table}
\caption{Interband gaps of C12A7. All values are in eV. The experimental (Exp.) and calculated (Calc.) gaps for sample A and sample B are included here, the type of sample is identified by the type of transition. The transitions measured in C12A7:e$^{-}$ via XPS by Toda~\cite{toda_07} and Souma,~\cite{souma_10} and the transitions measured in C12A7:O$^{2-}$ via optical spectroscopy by Hayashi~\cite{PVS_2007_JPCB_C12A7_OA_Oxygens} are included for comparison.}
\label{tbl:gaps}
\begin{tabular}{cccccc}
Transition & Exp. & Calc. & Toda & Souma & Hayashi\\
\hline
FVB-FCB (1a,c) & 5.9 & 6.0 & 7.5 & 7 & 6.6 \\
FVB-CCB (2c) & 5.0 & 4.1 & 5.5 & 5.2-6.2 & 5\\
FVB-O(2-) (3a) & 2.2 & 1.9/- & & & \\
O(2-)-FCB (4a) & 3.7 & 4.1/- & & & 4.9 \\
CCB-FCB (4c) & 1.4 & 1.6 & 2 & 2 &  \\
O(2-)-CCB (5a) & 2.8 & 2.2 & & & \\
\end{tabular}
\end{table}

To summarize, we have investigated the valence and conduction band structure of conducting and insulating C12A7 using \textit{ab initio} calculations and XPS, XES, and XAS measurements. The results of the theoretical calculations are in good quantitative agreement with the results of the spectroscopic measurements. We find clear experimental evidence of both the valence band states associated with extra-framework O$^{2-}$ ions and of the conduction band states associated with the  cage conduction band. Since the combination of XES/XPS and XAS measurements provides probes of the valence and conduction bands independently - while previous optical studies were confined to probe only transitions between the valence and conduction bands, and pervious XPS studies were confined to only probing the valence band - this study provides evidence in favour of the cage conduction band model for electrical conductivity in C12A7:e$^-$. Finally, the width of the core O \textit{1s} and Al \textit{2p} spectra in stoichiometric and oxygen deficient C12A7 provides evidence for the effect of the extra-framework O$^{2-}$ ions on the lattice structural disorder consistent with the earlier optical spectroscopy studies.~\cite{PVS_2007_JPCB_C12A7_OA_Oxygens}  C12A7:e$^-$ is one of the few stable electron-doped materials (``electrides'') and may well prove to be a useful material in many practical applications.

\begin{acknowledgments}
We gratefully acknowledge support from the Natural Sciences and Engineering Research Council of Canada (NSERC) and the Canada Research Chair program and JSPS FIRST Program. This work was done with partial support of the Russian Science Foundation for Basic Research (Project No. 11-02-00022). PVS is supported by the Royal Society. The calculations have been performed at the HECTOR facility (access provided via the Materials Chemistry Consortium, EPSRC grant EP/F067496) and at EMSL, a national scientific user facility sponsored by the Department of Energy's Office of Biological and Environmental Research and located at Pacific Northwest National Laboratory.
\end{acknowledgments}

\bibliography{cement}

\end{document}